\documentclass[12pt]{article}

\usepackage[dvips]{graphics}

\def\>{\rangle}

\def\<{\langle}

\newcommand{\rv}[1]{|#1\rangle}

\def\P{{\cal P}}

\newcommand{\nc}{\newcommand}

\nc{\figcap}[1]{\begin{quote}\refstepcounter{figure}

        {\bf Figure \thefigure}: {\small #1}\end{quote}}  
\nc{\fig}[1]{\mbox{Fig.~\ref{#1}}}

\nc{\noi}{\noindent}

\nc{\bea}{\begin{eqnarray}}  
\nc{\eea}{\end{eqnarray}}  
\nc{\bean}{\begin{eqnarray*}}  
\nc{\eean}{\end{eqnarray*}}  
\nc{\ba}{\begin{array}}  
\nc{\ea}{\end{array}}  
\nc{\be}{\begin{equation}}  
\nc{\ee}{\end{equation}}  
\nc{\nn}{\nonumber}  
\nc{\bra}[1]{\langle #1|}  
\nc{\ket}[1]{|#1\rangle}  
\nc{\av}[1] {\langle #1\rangle}  
\nc{\vac}[1] {\langle 0| #1|0\rangle}  
\nc{\amp}[2]{\langle #1|#2\rangle}  
\nc{\da}{\dagger}  
\nc{\pa}{\partial}  
\nc{\ga}{\gamma}  
\nc{\ep}{\epsilon}  
\nc{\tf}{t_f}  
\nc{\half}{\ensuremath{\frac{1}{2}}}  
\nc{\hHH}{\hat H}  
\nc{\ha}{\hat a}  
\nc{\hO}{\hat O}  
\nc{\hAA}{\hat A}  
\nc{\hB}{\hat B}  
\nc{\hG}{\hat G}  
\nc{\hN}{\hat N}  
\nc{\hU}{\hat U}  
\nc{\hx}{\hat{x}}  
\nc{\hp}{\hat{p}}  
\nc{\hpsi}{\hat \psi}  
\nc{\hphi}{\hat \phi}  
\nc{\hpi}{\hat \pi}  
\nc{\hpd}{\hat \psi ^\dagger}  
\nc{\hE}{\hat E}  
\nc{\hb}{\hat b}  
\nc{\hc}{\hat c}  
\nc{\hjo}{\hat j _0}  
\nc{\hrho}{\hat \rho}  
\nc{\leave}{\! \! \! \! \! / \, \,}  
\nc{\intl}[1]{\int d\! #1 \,} %defines a good spacing for integrals  
\nc{\intll}[3]{\int _#1^#2 d\! #3 \,} % integral with limits  
\nc{\lm}{\lim _{y \rightarrow x}}  
  
\nc{\scd}{\partial ^2 _{A_T}}  
\nc{\fd}[1]{\frac{\delta }{\delta #1}} % functional derivative  
\nc{\pad}[1]{\frac{\partial}{\partial #1}} % partial derivative  
\nc{\refpa}[1]{(\ref{#1})} % referencing equations with brackets  

\nc{\calH}{\ensuremath{\mathcal{H}}}  
\nc{\calD}{\ensuremath{\mathcal{D}}}  
\nc{\calL}{\ensuremath{\mathcal{L}}}  
\nc{\calO}{\ensuremath{\mathcal{O}}}  
\nc{\hcalO}{\ensuremath{\hat \mathcal{O}}}  
\nc{\calK}{\ensuremath{\mathcal{K}}}  
\nc{\Tr}{\ensuremath{\mathrm{Tr}}}  
\nc{\tr}{\ensuremath{\mathrm{tr}}}  
\nc{\ra}{\rightarrow}  
\nc{\lr}{\leftrightarrow}  
\nc{\phistar}{\phi^*}  
\nc{\etat}{\eta_T}  
\nc{\het}{\hat E_T}  
\nc{\hpt}{\hat \psi_T}  
\nc{\hpdt}{\hat \psi ^\dagger_T}  
\nc{\bart}{\bar{t}}  
\nc{\barp}{\bar{p}}  
\nc{\barT}{\bar{T}}  
\nc{\hbarrho}{\hat{\bar{\rho}}}  
\nc{\bga}{\ensuremath{\mbox{\boldmath{$\gamma$}}}}  
\nc{\bsi}{\ensuremath{\mathbf{\sigma}}}  
\nc{\bx}{\ensuremath{\mathbf{x}}}  
\nc{\by}{\ensuremath{\mathbf{y}}}  
\nc{\bz}{\ensuremath{\mathbf{z}}}  
\nc{\bp}{\ensuremath{\mathbf{p}}}  
\nc{\bn}{\ensuremath{\mathbf{n}}}  
\nc{\bbp}{\ensuremath{\bar{\mathbf{p}}}}  
\nc{\bP}{\ensuremath{\mathbf{P}}}  
\nc{\hbA}{\hat{\ensuremath{\mathbf{A}}}}  
\nc{\hbB}{\hat{\ensuremath{\mathbf{B}}}}  
\nc{\bA}{\ensuremath{\mathbf{A}}}  
\nc{\bJ}{\ensuremath{\mathbf{J}}}  
\nc{\bB}{\ensuremath{\mathbf{B}}}  
\nc{\bH}{\ensuremath{\mathbf{H}}}  
\nc{\bM}{\ensuremath{\mathbf{M}}}  
\nc{\bD}{\ensuremath{\mathbf{D}}}  
\nc{\bE}{\ensuremath{\mathbf{E}}}  
\nc{\hbE}{\hat{\ensuremath{\mathbf{E}}}}  
\nc{\br}{\ensuremath{\mathbf{r}}}  
\nc{\bj}{\ensuremath{\mathbf{j}}}  
\nc{\bOm}{\ensuremath{\mathbf{\Om}}}  
\nc{\om}{\omega}  
\nc{\Om}{\Omega}  
\nc{\sgn}{\mbox{sgn}}  
\nc{\deltabar}{\mbox{$\delta\hspace*{-8pt}\vspace*{-8pt}-$}}  
\nc{\gammat}{\tilde{\gamma}}  
\nc{\mub}{\bar{\mu}}  
\nc{\rhob}{\bar{\rho}}  
\nc{\Bb}{\bar{B}}  
\nc{\Jb}{\bar{J}}  
\nc{\Mb}{\bar{M}}  
\nc{\Tb}{\bar{T}}  
\nc{\sbar}{\bar{s}}  
\nc{\betab}{\bar{\beta}}  
\nc{\hj}{\hat j}  
\nc{\hQ}{\hat Q}  
\nc{\hJ}{\hat J}  
\nc{\hA}{\hat A}  
\nc{\hH}{\hat H}  
\nc{\de}{\delta}  
\nc{\leri}{\leftrightarrow}  
\nc{\llabel}[1]{\label{#1}\marginpar{#1}}

\newcommand{\bigpar}[1]{\ensuremath{\left#1 {\vrule height0.54em  
        width0em depth0.54em} \right.}}  
\newcommand{\wpp}[1]{\bra{\psi^{\pm}_{#1}}W(t)\ket{\psi^{\pm}_{#1}}}  
\newcommand{\wmm}[1]{\bra{\psi^{\mp}_{#1}}W(t)\ket{\psi^{\mp}_{#1}}}  
\newcommand{\wpm}[1]{\bra{\psi^{\pm}_{#1}}W(t)\ket{\psi^{\mp}_{#1}}}  
\newcommand{\wmp}[1]{\bra{\psi^{\mp}_{#1}}W(t)\ket{\psi^{\pm}_{#1}}}  
  
\newcommand{\wspp}[1]{\bra{\psi^{+}_{#1}}W(t)\ket{\psi^{+}_{#1}}}  
\newcommand{\wsmm}[1]{\bra{\psi^{-}_{#1}}W(t)\ket{\psi^{-}_{#1}}}  
\newcommand{\wspm}[1]{\bra{\psi^{+}_{#1}}W(t)\ket{\psi^{-}_{#1}}}

\newcommand{\ex}[2]{e^{#1 igt\sqrt{#2}}}  
\newcommand{\exx}[4]{e^{#1 igt\left( \sqrt{#2} #3 \sqrt{#4}\right)}}  
  
\nc{\bc}{\begin{center}}  
\nc{\ec}{\end{center}}  
\nc{\inv}[1]{\frac{1}{#1}}

\newlength{\overeqskip}  
\newlength{\undereqskip}  
\nc{\eq}[1]{\mbox{Eq.~(\ref{#1})}}  
\nc{\eps}{\epsilon}  
\nc{\goto}{\rightarrow}

\nc{\cF}{{\cal F}}  
\nc{\cG}{{\cal G}}  
\nc{\cH}{{\cal H}}

\begin{document}  
%%%%%%%%%%%%%%%%%%%%%%%%%%%%%%%%%%%%%%%%%%%%%%%%%%%%%%%%%%%%%%%%%%%  
%\documeinstantaneouslyntclass[12pt]{article}  
%  
%\begin{document}   
%   
%   
%  
%   
%   
%   
\thispagestyle{empty}   
%   
%  
%\begin{flushright}{\begin{tabular}{l}   
%  
%  
{\begin{tabular}{lr}   
%  
%   
%\today    
& ~~~~~~~~~~~~~~~~~~  
~~~~~~~~~~~~~~~~~~~~~~~~~~~~~~~~~~~~~~~~~~~~~~~~~~~~~~~~~~~CERN-TH/99-412  
\end{tabular}}   
%   
%\end{flushright}   
%   
\vspace{10mm}   
\begin{center}   
\baselineskip 1.2cm   
{\Huge\bf   Macroscopic Interference Effects in Resonant Cavities  
}\\[1mm]   
\normalsize   
\end{center}   
{\centering   
{\large Bo-Sture K. Skagerstam\footnote{Emil   
address: boskag@phys.ntnu.no.}$^{,a,b}$}   
{\large Bj\o rn \AA .  
Bergsjordet\footnote{Email   
address:  bjorberg@phys.ntnu.no.}$^{,a}$}   

{\large Per K. Rekdal\footnote{Email address: perr@phys.ntnu.no.}$^{,a}$   
\\[5mm]   
  
$^{a}~$Department of Physics,   
The Norwegian University of Science and Technology,     
N-7491 Trondheim, Norway \\[1mm]  
$^{b}~$CERN, TH-Division, CH-1211 Geneva 23, Switzerland\\[1mm]  
} }   
%   
%[-5mm]}}   
%   
%   
%   
\begin{abstract}   
\normalsize   
%   
%\vspace{-5mm}   
%   
%\vspace{5mm}   
%   
\noindent   
We investigate the possibility of interference effects induced by   
a macroscopic  
quantum-mechanical superposition of almost orthogonal coherent states - a Schr\"{o}dinger  
cat state - in a resonant microcavity. Despite the fact that a single  
atom, used as a probe of the cat state, on the average only change  
the mean number of photons by one unit, we show that this single atom  
can change the system drastically. Interference between the initial  
and almost orthogonal macroscopic quantum states of the radiation field can  
now take place. Dissipation under current experimental conditions is  
taken into account and it is found that this does not 
necessarily change the  
interference effects dramatically.       
\end{abstract}   
\newpage  
%=============================================================   
%   
\vspace{1cm}  
%%%%%%%%%%%%%%%%%%%%%%%%%%%%%%%%%%%%%%%%%  
\bc{  
\section{INTRODUCTION}  
}\ec  
%%%%%%%%%%%%%%%%%%%%%%%%%%%%%%%%%%%%%%%%%  
\vspace{0.5cm}  
  
Superpositions of orthogonal states which exhibits macroscopic  
features appear to be of fundamental importance in recent studies of the  
foundations of quantum mechanics (see e.g.  
Ref.\cite{gimo&97,knight&95} and references cited therein).  As was  
noticed by Schr\"{o}dinger \cite{schrod&35}, a direct extrapolation of  
the calculational rules of quantum mechanics to macroscopic systems  
would lead to the appearance of quantum-mechanical interference  
effects of classical objects like a living and a dead organism. We  
have overwhelming empirical evidence that such macroscopic  
interferences are rare or absent in the real world. It can actually be argued that such quantum mechanical macroscopic superpositions decay with a very  
short life-time due to environment-induced decoherence \cite{decoherence}.  
  
Resonant microcavities can be used to study the behaviour of  
mesoscopic superpositions coherent states.  In order to be specific we  
will, as an initial state of the cavity radiation field, consider a superposition of coherent  
states as an example of a Schr\"{o}dinger "cat" state.   
The non-dissipative dynamics of the atom-photon interaction, to be described below, is assumed to be  
described by the Jaynes-Cummings (JC) model \cite{Jaynes63}. In terms  
of conventional coherent states $\rv{z}$   
parameterised in terms of a complex number $z$ (see e.g.  Refs.\cite{Klauder&Skagerstam&85}), this Schr\"{o}dinger  
cat state has the following form  
\begin{equation}  
\label{eq:catstate}  
\rv{z;\phi} = \frac{1}{(2+2\cos\phi\exp(-2|z|^2))^{1/2}} (\rv{z} +  
e^{i\phi}\rv{-z})~~~.  
\end{equation}  
Such states have been studied in great detail in the literature  
(see e.g. Refs.\cite{knight&95}, \cite{puri&86}-\cite{pyatos&96}). It can be shown that  
the results presented below do not  
crucially depend of the form of the initial cavity state. The  
essential ingredient is that the overlap probability $|\amp{z}{-z}|^2 = \exp(-2|z|^2)$ is a sufficiently small number. In our case we will for example consider  
examples with an average number of photons $\bar{n}=|z|^2=49$ and therefore $|\amp{z}{-z}|^2 \approx 10^{-43}$, indeed a very small number.

The micromaser system \cite{Filipowicz86} is an experimental realisation of the idealised  
system of a two-level atom interacting with a second quantized  
single-mode of the electro-magnetic field (for reviews and references see  
e.g.  \cite{Walther88}). The microlaser \cite{An94} is  
the counterpart in the optical regime.  Trapping states \cite{Fili&Java&Meystre}   
 - \cite{Slosser&Meystre_II} have recently  
been generated in the stationary state of the micromaser system and therefore the generation  
of Schr\"{o}dinger cat states in such a system may be feasible \cite{trapping&99}. In resonant cavities  
other possibilities also exist \cite{brune&96} in which case  
decoherence actually has been studied experimentally. Schr\"{o}dinger  
cat states have also been studied experimentally in other systems like in atomic systems \cite{stroud&96}, and in ion-traps \cite{meekhof&96}. Related revival phenomena has also been  
studied for Bose-Einstein condensates \cite{pita&98}.  
  
\begin{figure}[tp]   
\unitlength=1mm  
\begin{picture}(200,170)(-15,-5)  
\includegraphics{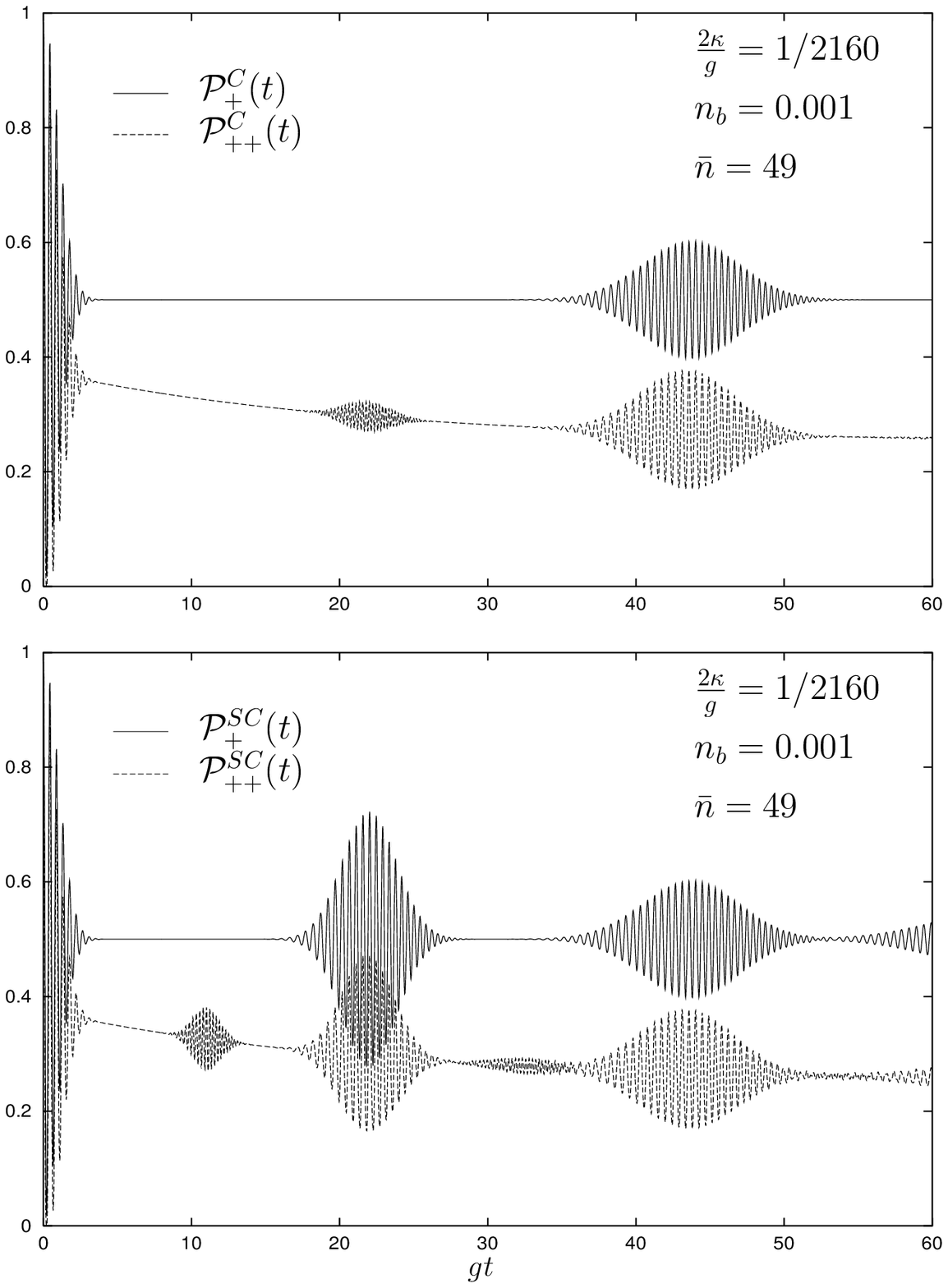}  
\end{picture}  
\figcap{The upper figure shows the revival probabilities  
 $\P_{+}(t)$ and $\P_{++}(t)$ for ($C$) a  coherent state $\rv{z}$  with a mean number  
 $\bar{n}_C =|z|^2 =49$ of photons as a  
 function of the atomic passage time $gt $.  
The lower figure shows the same revival probabilities for ($SC$) an even, i.e. $\phi =0$,  
Schr\"{o}dinger cat state $\rv{z;\phi}$  with a mean-value of $\bar{n}_{SC} = |z|^2 \tanh(|z|^2)  \approx 49$  of  
photons. Here we have used physical parameters $\kappa = 8.33\ s^{-1}$ and a  
Rabi frequency $g=36\ kHz$ corresponding to the parameters of  
 Ref.\cite{benson&97}.  
\label{fig:JC1revivals}}  
\end{figure}

We assume that the atoms which enter the microcavity one at a time,  are all prepared in the excited state. Each atom spends a time $t$ in the microcavity interacting with the radiation field. It then leaves the microcavity and the state of the atom is measured.  
Let ${\cal P}_{s_{1}}(t)$ be the probability that an atom is found in the  state $s_{1}=\pm$, where $+(-)$ denotes the excited(ground) state, after it leaves the microcavity. Similarly, let ${\cal P}_{s_{1}s_{2}}(t)$ be the probability that the next atom is in the state $s_{2}=\pm$ if the previous atom  has been found in the state $s_{1}$. ${\cal P}_{s_{1}}(t)$ exhibits well known revivals (see e.g. Refs.\cite{Averbukh&91}), which has been observed experimentally in microcavity systems \cite{RWK_87,BSMDHRH_96} and in ion-traps \cite{meekhof&96}. ${\cal P}_{s_{1}s_{2}}(t)$ exhibits in addition so called pre-revivals \cite{ElmforsLS95}. In the upper figure of Fig.\ref{fig:JC1revivals} such revivals ($gt_{rev}\approx 2\pi\sqrt{\bar{n}}$) and prerevivals ($gt_{rev}\approx \pi\sqrt{\bar{n}}$) are illustrated for a mesoscopic coherent state with $\bar{n}=49$, using  the physical cavity parameters of Ref.\cite{benson&97}. Damping effects at a non-zero temperature, to be discussed below,  are included in Fig.\ref{fig:JC1revivals}. In the lower figure of Fig.\ref{fig:JC1revivals} we  
exhibit the same probabilities but for an even, i.e. $\phi = 0$, Schr\"{o}dinger cat state with $\bar{n}\approx 49$ and we observe new clear revival phenomena.   
  
It is possible to give a simple physical explanation of these additional revivals, which occur approximately at half of the revival and prerevival times of a coherent state with the same mean-value of photons \cite{seoul&99}.   
It is known that the JC-model has the property that with an initial coherent state of the radiation field, one reaches a pure atomic state at $gt \approx gt_{rev}/2$,  independent of the initial atomic state \cite{banac90}. The coherent states  $\rv{z}$ and $\rv{-z}$ used in the construction of the  
Schr\"{o}dinger cat state Eq.(\ref{eq:catstate})  are approximately orthogonal. These states will then, as independent states, lead to the same atomic state at $gt \approx gt_{rev}/2$ up to a phase and a quantum-mechanical interference pattern will emerge due to the evolution of different "paths" to the same final state. The states $\rv{z}$ and $\rv{-z}$ of the radiation field therefore, in a sense , describe an atomic interferometer. As we will argue below, this interferometer is actually quantum-mechanical since the interference pattern depends on the relative phase $\phi$ of the Schr\"{o}dinger cat state Eq.(\ref{eq:catstate}).   
  
 We therefore have the interesting situation of  
a superposition of "macroscopic" states, which can be interpreted as a classical interferometer for atoms, and where, in addition, the different "classical slits" of the interferometer  interfere quantum-mechanically. With an increasing number of photons in the cavity this interference vanishes rapidly.  
In fact, if $t_{cav}$ is the decay time of the cavity, decoherence effects will be operative on a time-scale   $t_{d}\approx t_{cav}/\bar{n}(1+n_b)$, which for $n_b = 0$ agrees with a known result \cite{walls&85}. If $\bar{n}$ is to small revival phenomena will, on the other hand, not be very pronounced.  We will argue below that for  
a realistic experimental situation one can choose a time scale $t$, such that $t\ll t_d$, and an average  number of photons $\bar{n}$ for which the interference effect above  
 still should be realisable in the laboratory.   
  
The present paper is organised as follows. In Section \ref{sec:dyn} we recapitulate, for the convenience of the reader and in order to define the notation used in the present paper, the basic ingredients in the JC-model. The effect of cavity damping on ${\cal P}_{s_1}(t)$ and ${\cal P}_{s_1s_2}(t)$ is discussed in Section \ref{sec:inter} extending the analysis of Refs.\cite{puri&86}-\cite{barnett&86} to  a small and non-zero $n_b$. In Section \ref{sec:final} we summarise our results together with some final remarks. Some equations used in the main text are  
summarised in an Appendix.

\vspace{1cm}   
%%%%%%%%%%%%%%%%%%%%%%%%%%%%%%%%%%%%%%%%%  
\bc{  
\section{THE DYNAMICAL SYSTEM}  
\label{sec:dyn}  
}  
\ec  
%%%%%%%%%%%%%%%%%%%%%%%%%%%%%%%%%%%%%%%%%  
%\vspace{0.5cm}  
  
The electro-magnetic interaction between a two-level atom, with level  
separation $\omega_0$, and a single mode  of the  
radiation field in a cavity with frequency $\omega$ is described, in the rotating wave  
approximation, by the JC Hamiltonian  
\cite{Jaynes63}  
\begin{equation}  
\label{eq:JCH}  
H=\omega a^{*} a+\frac12\omega_0\sigma_z+g(a\sigma_++a^*\sigma_-)~~,  
\end{equation}  
\noi where the coupling constant $g$ is proportional to the dipole  
matrix element of the atomic transition.  
Here we make use of the Pauli matrices to describe the two-level atom and the  
notation $\sigma_\pm=(\sigma_x\pm i\sigma_y)/2$.  The second-quantized  
single mode electro-magnetic field is described in a conventional  
manner by means of an annihilation (creation) operator $a$ ($a^*$),  
where we have suppressed the cavity mode quantum numbers. For $g=0$ the  
atom-field states $|n,s\>=|n\>\otimes|s\>$ are characterised by  
the quantum number $n=0,1,\ldots$ of the oscillator and $s=\pm$ for  
the atomic levels with energies $  
E_{n,\pm}= \omega n \pm \omega_0 /2$.  
At resonance $\omega=\omega_0$ the levels $|n-1,+\>$ and $|n,-\>$ are  
degenerate for $n\ge1$ (except for the ground state $n=0$), but this  
degeneracy is lifted by the interaction.  For an arbitrary coupling $g$  
and detuning parameter $\Delta\omega= \omega_0-\omega$ the system  
reduces to a $2\times2$ eigenvalue problem, which may be trivially  
solved \cite{Jaynes63}. The result, which describe the entangled system of an atom and the radiation field, is that two new dressed levels, $|n,1\>$ and $|n,2\>$, are  
formed as superpositions of the previously degenerate ones at resonance according to  
\begin{eqnarray}  
\label{eq:JStates}  
|\psi^{+}_{n}\>& =& \cos\theta_n|n+1,-\> +\sin\theta_n|n,+\>~~,\\  
|\psi^{-}_{n}\>& =& -\sin\theta_n|n+1,-\> +\cos\theta_n|n,+\>~~,  
\end{eqnarray}  
\noindent with energies   
\begin{eqnarray}  
\label{eq:Jenergies}  
E_{n}^+ = \omega(n+1/2) + \sqrt{\Delta\omega^2/4+g^2(n+1)}~~,\\  
E_{n}^- = \omega(n+1/2) - \sqrt{\Delta\omega^2/4+g^2(n+1)}~~,  
\end{eqnarray}  
respectively.  The ground-state of the coupled system is given by  
$|\psi_0\> = |0,-\>$ with energy $E_0 = -\omega_0 /2$. Here the mixing angle  
$\theta_n$ is given by  
\begin{equation}  
\label{eq:Jmixing}  
\tan\theta_n = \frac{2g\sqrt{n+1}}{\Delta\omega  +  
  \sqrt{\Delta\omega^2 + 4g^2(n+1)}}~~.  
\end{equation}  
%  
%  
%\noindent   
The interaction therefore leads to a separation in energy $\Delta E_n  
= \sqrt{\Delta\omega^2+4g^2(n+1)}$ for the quantum number $n$.  The system  
performs Rabi oscillations with the corresponding frequency between  
the original, unperturbed states.  The two-level atoms which  
enter the cavity are assumed to be prepared in the excited state, i.e.  
the density matrix is of the diagonal form  
  
\begin{equation} \label{pho_A}   
   \rho_A =   
   \left  (  
          \ba{rr}  
           1 & 0  \\  
           0 & 0  
          \ea   
   \right )~~,  
\end{equation}  
The initial density matrix $\rho_C$ of the cavity radiation field is  
determined by the Schr\"{o}dinger cat state Eq.(\ref{eq:catstate}).  
  
The damping of the cavity is described by a conventional master  
equation (see e.g. Refs.\cite{Agarwal73,Walls95}), i.e. if $\rho $ is  
the density matrix of the combined atom-field system we have that   
\begin{equation}  
\label{eq:Damping}  
{\displaystyle\frac{d\rho}{dt}}=~~i[\rho,H]   
-\kappa(n_b+1)(a^*a\rho +\rho a^*a-2a\rho a^*)   
-\kappa n_b(aa^*\rho +\rho aa^*-2a^*\rho a)~~,  
\end{equation}  
%  
%\noi   
where $n_b$ is the average occupation number of thermalized cavity photons at the  
oscillator frequency and $2\kappa = 1/t_{cav}$ is the decay constant of the  
cavity.

\vspace{1cm}   
%%%%%%%%%%%%%%%%%%%%%%%%%%%%%%%%%%%%%%%%%  
\bc{   
  \section{INTERFERENCE EFFECTS}  
\label{sec:inter}  
}\ec  
%%%%%%%%%%%%%%%%%%%%%%%%%%%%%%%%%%%%%%%%%  
%\vspace{1cm}  

We now define  
%  
%\begin{align}  
\begin{equation}  
  W(t) = e^{iHt}\rho (t) e^{-iHt}\ ,\label{Wdef}  
%\end{align}  
\end{equation}  
which allows us to write the equations for the diagonal elements of  
Eq.~(\ref{eq:Damping}) in the form given in  
the Appendix.  By introducing  
%  
%\begin{align}  
\begin{equation}  
  F_n  \equiv F_{n}(t) = \langle  
    \psi^{+}_{n}|W(t)|\psi^{+}_{n}\rangle + \langle   
    \psi^{-}_{n}|W(t)|\psi^{-}_{n}\rangle \ ,   
\end{equation}  
  for $n\geq 0$ and the ground-state expectation value  
\begin{equation}  
  F_{-1}  \equiv F_{-1}(t) = 2\langle{\psi_0 |W(t)|}\psi_0 \rangle \ ,  
\end{equation}  
%\end{align}  
%  
and by omitting rapidly oscillating terms valid under the assumption $g \ll \kappa$ \cite{puri&86,barnett&86}, the equations of motion can be considerably simplified.  
We obtain  
\begin{equation}  
  \dot{F}_{n}\equiv \frac{dF_{n}(t)}{dt} = -\alpha_{n} F_{n}+\beta_{n} F_{n+1} +\gamma_{n}  
  F_{n-1} \ , \label{diff2} \\   
\end{equation}  
  where we have defined  
 \begin{eqnarray}  
    \alpha_{n} &=& 2\kappa \bigpar{(}2n_b(n+1) +   
    n+\frac{1}{2}\bigpar{)}  \ ,\nonumber \\  
    \beta_{n} &=& 2\kappa (n_b +1)(n+\frac{3}{2}) \ ,\\  
    \gamma_{n} &=& 2\kappa n_b (n+\frac{1}{2})  \ , \nonumber  
  \end{eqnarray}  
  for $n \geq 0$, and where  
 \begin{eqnarray}  
    \alpha_{-1} &=& 2\kappa n_b  \ ,\nonumber\\  
    \beta_{-1} &=&  2\kappa (n_b +1 ) \ ,\\  
    \gamma_{-1} &=&  0 \ . \nonumber  
 \end{eqnarray}  
%\end{align}  
%  
We observe that $\alpha_{n}- \beta_{n-1}-\gamma_{n+1}=0$ if $n\geq 1$ and   
$\alpha_{0}- \beta_{-1}-\gamma_{1}=-\beta_{-1}/2$. In order to find an approximate solution $F^{*}_{N}$ of  
Eq.~(\ref{diff2}), valid for sufficiently small $n_b$, we proceed as follows. For sufficiently large but finite $n=N$, we put $F^{*}_{N+1} = 0$ as in Ref.\cite{puri&86}.  In addition we make use of the  
approximation \( \gamma_n F^{*}_{n-1} \simeq \gamma_n F^{*}_{n} \mbox{  
  for all } n \leq N \) .  With these approximations the equation of  
motion for $n=N$ is  
 \begin{equation}  
  \dot{F}^{*}_N  = (-\alpha_N +\gamma_N)F^{*}_N \equiv -\alpha'_N F^{*}_N  \ ,  
 \end{equation}  
with the simple solution  
 \begin{equation}  
F^{*}_N(t) = F^{*}_N (0) \, e^{-\alpha'_N t} \label{fn}\ .  
 \end{equation}  
Next, we solve the equation  
 \begin{equation}  
    \dot{F}^{*}_{N-1}=-\alpha'_{N-1} F^{*}_{N-1}+\beta_{N-1}F^{*}_N \ ,   
 \end{equation}  
i.e. by making use of Eq.(\ref{fn})  
 \begin{equation}  
  \ddot{F}^{*}_{N-1} +\alpha'_{N-1} \dot{F}^{*}_{N-1}   
% &= \beta_{N-1}  
%   \dot{F}^{*}_N \nonumber \\  
   = -\beta_{N-1}\alpha'_N e^{-\alpha'_{N-1}t}F^{*}_N (0) \ .  
 \end{equation}  
As one easily can verify, the  
solution of this equation is  
 \begin{equation}  
  F^{*}_{N-1}  =  e^{-\alpha'_{N-1}t} \bigpar{[} F^{*}_{N-1}(0)+  
  \frac{\beta_{N-1}}{\alpha'_{N-1}-\alpha'_{N}}\bigpar{(}   
  e^{(\alpha'_{N-1}-\alpha'_N ) t} - 1\bigpar{)} F^{*}_N (0)\bigpar{]} \ .  
 \end{equation}  
Iteration of the previous procedure leads to the approximative solution  
 \begin{equation}  
  F^{*}_n  = e^{-2\kappa t\left[ (n+\frac{1}{2}) (n_{b} +1) +n_{b}  
  \right]}   
  \sum_{j=n}^{N} \frac{(j+\frac{1}{2})!\left( 1-  
  e^{-2\kappa t(n_b+1)} \right)^{j-n}} {(n+\frac{1}{2})!(j-n)!}p_{j}  
  \label{fn2}\ ,   
 \end{equation}  
valid for $n\geq0$.  Here $p_{j}$ is the initial photon probability distribution  
of the cavity radiation field.  
The approximative solution $F^{*}_{n}$ deviates from the exact solution $F_{n}$ in a manner  
which can  be expressed in terms of the right-hand side in the equation  
 \begin{equation}  
  \dot{F}^{*}_n +\alpha_n F^{*}_n -\beta_n F^{*}_{n+1} -\gamma_n F^{*}_{n-1}  =   
  \gamma_n (F^{*}_n -F^{*}_{n-1}) \ , \label{feil}  
 \end{equation}  
by comparing with Eq.~(\ref{diff2}).  
By performing a sum on both sides of  
Eq.(\ref{feil}) we now obtain  
 \begin{equation}  
\label{middle}  
  \beta_{-1} F^{*}_0+ \sum_{n=0}^{N} \bigpar{[}  \dot{F}^{*}_n +\alpha_n F^{*}_n  
  -\beta_{n-1} F^{*}_{n} -\gamma_{n+1} F^{*}_{n}\bigpar{]}   
   = -\alpha_{-1} \sum_{n=0}^{N}F^{*}_n \label{feil2} \ .  
 \end{equation}  
By making use of the unitarity relation  
\(\mbox{Tr}(\rho(t))=\sum\limits_n p_n=1\), we can therefore find an explicit expression for  
  $F^{*}_{-1}$, i.e.  
 \begin{equation}  
\label{fminus1}  
  F^{*}_{-1} = 2\left(1-\sum_{n=0}^{N} F^{*}_n \right) =  
  2 - e^{-2\kappa n_{b} t}   
  \sum_{j=0}^{N}\sum_{k=0}^{j}(-1)^{k} \frac{(j+\frac{1}{2})!} {(j-k)!k!(\frac{1}{2})!}\frac{e^{-\kappa(2k+1)(n_b+1)t}}{k+\frac{1}{2}}~p_{j}\ , \end{equation}  
such that $F^{*}_{-1}(0)=0$. Here we observe  that $F^{*}_{-1}$  satisfies the following differential equation   
 \begin{equation}  
  \dot{F}^{*}_{-1}  + \alpha_{-1}F^{*}_{-1} -\beta_{-1}F^{*}_0 =  
  4\kappa n_b \ ,   
  \label{feil3}     
 \end{equation}  
by making use of   
Eq.(\ref{middle}). From these considerations we conclude that, when $\gamma_n = n_b =0$,  
Eqs.~(\ref{fn2})~and~(\ref{fminus1}) give the exact solution to  
the equations of motion.  With our  
method we have therefore found an approximative solution at a non-zero $n_b$   
which is consistent with unitarity,  
as is required in quantum mechanics, and, furthermore, with a small error when $n_b$ is sufficiently small.  For the numerical results of the  
present paper we have indeed verified that the corresponding error can be neglected, at least  
under the condition as given in the end of the Appendix.  
In order to proceed further, we also need the result  
 \begin{equation}  
  \langle \psi^{\pm}_n |W(t)|\psi^{\mp}_n\rangle = \frac{1}{2}  
  e^{- \alpha_n t} p_n \ , \label{something}  
 \end{equation}  
which is easily found from the equations of motion in  
the Appendix. From the explicit expressions for  
$F^{*}_{n}$, $F^{*}_{-1}$ and Eq.(\ref{something}), which are the major results of the present paper,  we  see that $t_d \simeq t_{cav}/\bar{n}(n_b+1)$, at least for sufficiently small $n_b$.  
  
Using the Eqs.~(\ref{fn2}),~(\ref{fminus1})~and~(\ref{something}) we are now able to evaluate the time evolution of various physical quantities.   
If for example the first atom is measured to be in the excited state at $t=t_A$,  
the reduced density of the cavity radiation field is $\rho_\gamma (t_A) =  
\mbox{Tr}_A (\rho(t_A)|+\rangle\langle+|) = \langle  
+|\rho(t_A)|+\rangle$. Within our approximations we then obtain  
 \begin{equation}  
  p_n(t_A)=\langle n| \rho_{\gamma} (t_A )|n \rangle  = \frac{1}{2}  
  \bigpar{[}   
  F^{*}_n (t_A ) + e^{-\alpha_n t_A } \cos (2gt_A \sqrt{n+1}) ~p_n  
  \bigpar{]} \ .   
 \end{equation}  
The probability of finding the atom in an excited state at   
time $t$ then is  
 \begin{equation}  
  {\cal P}_+(t) = \frac{1}{2} - \frac{1}{4}F_{-1}^{*} +  
\sum_{n=0}^{N}  
\frac{1}{2} \left[ e^{-\alpha_{n}t} \cos \left( 2gt\sqrt{n+1}\right) p_n \right]  
 \ .  
\label{pplus}  
 \end{equation}  
For the next atom which enters into the cavity, $\rho_\gamma(t_A)$ is then used as the initial  
density of the radiation field.  The probability of finding the next  
atom in the excited state at time $t_B$, given that the first atom was  
in the excited state at time $t_A$, therefore is  
 \begin{eqnarray}  
 {\cal P}_{++}(t_B)   
&=& \sum_{n=0}^{N}\frac{1}{2}\bigpar{[}  
  e^{-2\kappa \left[ (n+\frac{1}{2}) (n_{b} +1) +n_{b}  
    \right] (t_B - t_A)} \nonumber \\  
  &\times& \sum_{j=n}^{N} \frac{(j+\frac{1}{2})!\left( 1-  
      e^{-2\kappa(n_b+1)(t_B -t_A)} \right)^{j-n}} {(n+\frac{1}{2})!(j-n)!}  
  ~p_j(t_A)  
  \nonumber\\   
  &+& e^{-\alpha_n (t_B -t_A)}\cos(2g(t_B -t_A)\sqrt{n+1})~  
  p_n(t_A)\bigpar{]} \ .  
  \label{pplusplus}   
 \end{eqnarray}  
From now on we  choose the atomic passage times $t_A$ and $t_B$ such that $t \equiv t_A = t_B - t_A$.   
% The numerical results for coherent- and Scr\"odinger cat states are  
% displayed in Figure~\ref{fig:JC1revivals}.  

\begin{figure}[t]   
\unitlength=1mm  
\begin{picture}(110,80)(0,0)  
\includegraphics{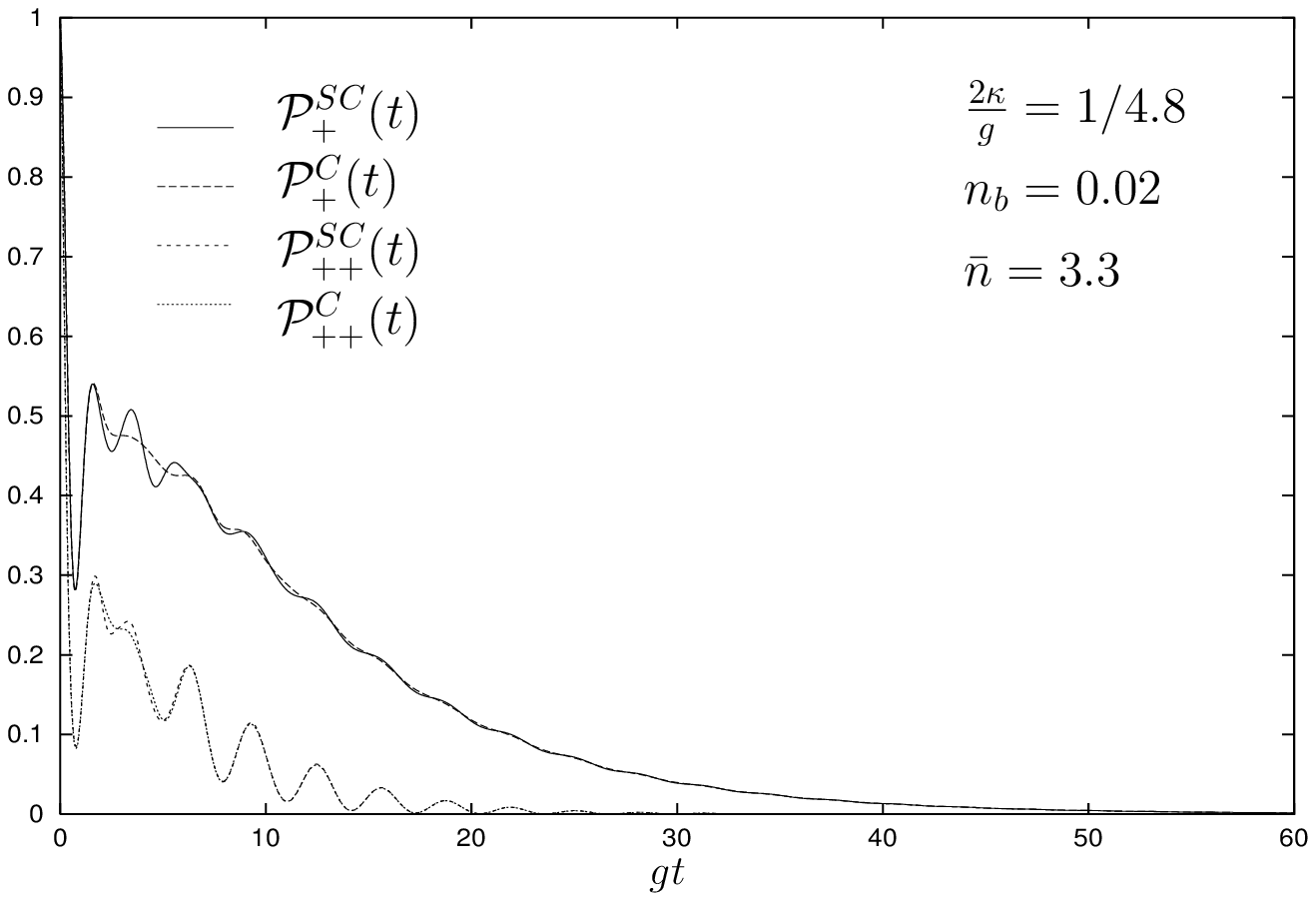}  
\end{picture}  
\figcap{The figure shows the revival probabilities $\P_{+}(t)$ (upper curves) and  
  $\P_{++}(t)$ (lower curves) for ($C$) a coherent state  $\rv{z}$  and ($SC$)  
an even Scr\"odinger cat state  $\rv{z,\phi=0}$, with a mean number $\bar{n}_C=|z|^2 =3.3$ and   
$\bar{n}_{SC}=|z|^2\tanh(|z|^2) \approx 3.3$, respectively, of photons as a function of  
  the atomic passage time $gt $.  Here we have used physical  
  parameters $\kappa = 2500\ s^{-1}$ and a Rabi frequency $g=24\ kHz$  
  corresponding to the parameters of Ref.\cite{brune&96}.  
  \label{fig:JC2revivals}}  
\end{figure}  
  
\begin{figure}[tp]   
\unitlength=1mm  
\begin{picture}(200,150)(-15,-5)  
\includegraphics{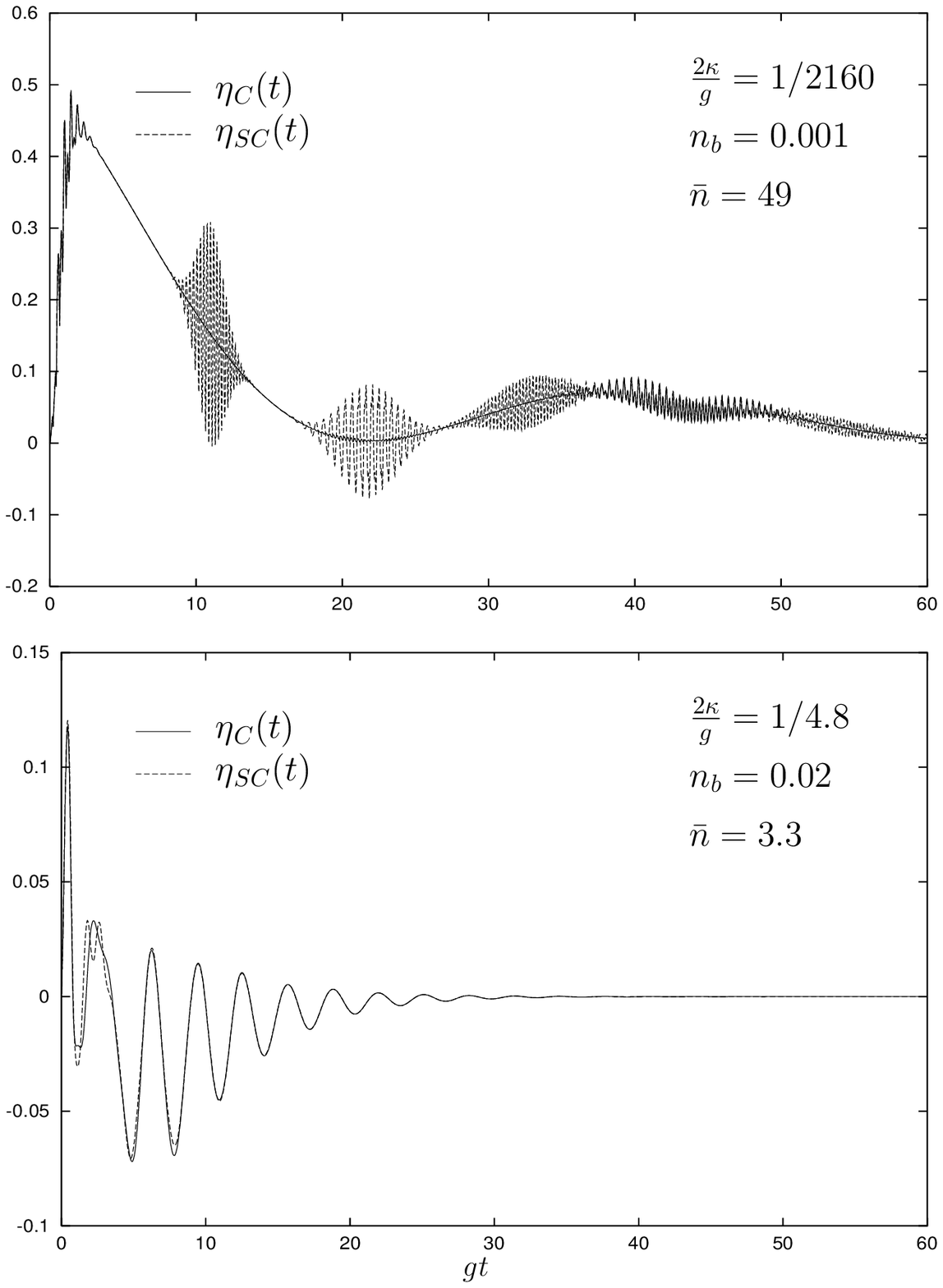}  
\end{picture}  
\figcap{The upper figure shows the correlation function $\eta$ as a  
  function of the atomic passage time for ($C$) a coherent state $\rv{z}$ and  
for ($SC$) a Schr\"odinger cat state $\rv{z,\phi=0}$ with the experimental parameters  
of Ref.\cite{benson&97} apart from our choice $\bar{n} = 49$. The lower figure shows the same correlation functions, but with the  
parameters of Ref.\cite{brune&96}.  
\label{fig:JC3revivals}}  
\end{figure}  
  
\newpage  
\vspace{1cm}  
%%%%%%%%%%%%%%%%%%%%%%%%%%%%%%%%%%%%%%%%%  
\bc{  
\section{FINAL REMARKS}  
\label{sec:final}  
}\ec  
%%%%%%%%%%%%%%%%%%%%%%%%%%%%%%%%%%%%%%%%%  
%\vspace{0.5cm}  
%  
  
In Fig.~\ref{fig:JC1revivals} we show the revival probabilities  
${\cal P}_{+}(t)$ and ${\cal P}_{++}(t)$ for an initial coherent state or a  
Schr\"odinger cat state, where the physical parameters are taken from  
 Ref.\cite{benson&97}. We notice that when the system initially is  
prepared in a Schr\"odinger cat state we observe a revival at an earlier  
time as compared to the case of an initial a coherent state. The  
first revival time for Schr\"odinger cat state is  
$t_{rev}^{SC}=t_{rev}^{C}/2$, where $t_{rev}^{C}$ is the first revival  
time for the coherent state.  
Similarly, in Fig.~\ref{fig:JC2revivals}, we show the same revival  
probabilities, but now the parameters are taken from Ref.\cite{brune&96}.  
Here we see only a minor difference between an initial Scr\"odinger cat state   
as compared to an initial coherent state.    
In Fig.~\ref{fig:JC3revivals} we have, in addition, evaluated the correlation function $\eta (t)$  defined by \cite{brune&96}  
 \begin{equation}  
\eta (t) = \frac{{\cal P}_{++}(t)}{{\cal P}_{+}(t)}-  
\frac{{\cal P}_{-+}(t)}{{\cal P}_{-}(t)} ,  
 \end{equation}  
such that $-1 \leq \eta (t) \leq 1$, for an initial coherent and Schr\"odinger cat state  
respectively. We notice that ${\cal P}_{-+}(t)+   
{\cal P}_{--}(t) = {\cal P}_{-}(t)= 1- {\cal P}_{+}(t)$. Again we notice that  
under the experimental conditions of Ref.\cite{benson&97}, with  a finite but not to large $\bar{n}$,   
one is able to very clearly detect the revivals of a Schr\"odinger cat state at $t_{rev}^{SC}$ by measuring the correlation function $\eta (t)$.  
  
The interference effect we have discussed above depends on the phase $\phi$   
of the Scr\"odinger cat state. One interesting case occurs when $\phi=\pi/2 ~\mbox{mod}(\pi)$. With regard to revivals, the  
Scr\"odinger cat state then behaves like a coherent state, i.e. the  
revival at $t^{rev}_{SC}$ is now absent. In fact, the phase $\phi$ essentially only effects  
the first revival at $gt_{rev}\approx \pi\sqrt{\bar{n}}$.  
This is easily realised in terms of a Poisson resummation of Eq.(\ref{pplus}) \cite{Poisson} for $\bar{n}\gg 1$,  i.e.   
 \begin{eqnarray}  
\label{Poisson}  
  {\cal P}^{SC}_{+}(t)  \approx  \frac{1}{2}e^{{\displaystyle -2\kappa n_{b} t}}  
~~~~~~~~~~~~~~~~~~~~~~~~~~~~~~~~~~~~~~\nonumber \\+~\frac{1}{2}e^{{\displaystyle -2\kappa t(2n_b(1+\bar{n})+\bar{n}+1/2)}} \left[ w_0(t)    + \sum_{\nu=1}^N\bigpar{(}w_{\nu}(t)-w_{\nu-1/2}(t)\cos\phi\bigpar{)}\right] \  .  
 \end{eqnarray}  
Here   
 \begin{equation}  
  w_0(t) =  
    e^{-{\displaystyle g^2t^2 / 2}} \cos\left(2gt\sqrt{\bar{n}}\right) \ ,  
 \end{equation}  
describes the initial exponential decay and the various revivals are expressed in terms of  
 \begin{equation}  
  w_{\nu}(t) = p(\frac{g^2t^2}{4\pi^2\nu^2})\frac{gt}{\pi\sqrt{2\nu^3}} \cos\left( \frac{g^2t^2}{2\pi\nu} -\frac{\pi}{4}\right) \ ,  
 \end{equation}  
where, in our case,  
 \begin{equation}  
p(n) = \frac{\bar{n}^{\displaystyle n}}{n!}e^{{\displaystyle -\bar{n}}} \ .  
 \end{equation}  
 In Eq.(\ref{Poisson}) only  a leading effect of damping is taken into account  
in that $F_{-1}^{*}\approx 2(1- e^{\displaystyle -2\kappa n_b t})$, vaild for $\alpha_{\bar{n}}\ll g$.   
Numerically Eq.(\ref{Poisson}) with $N=3$ is e.g. sufficient to describe  ${\cal P}^{SC}_{+}(t)$ of Fig.\ref{fig:JC1revivals}   
very accurately.  
  
We therefore conclude that the two, almost orthogonal, coherent states $\rv{z}$ and $\rv{-z}$ of the Scr\"odinger cat state Eq.(\ref{eq:catstate}), both with a large average number of photons, can act like an "interferometer" for the atoms with regard to the revival at, e.g., $t=t^{rev}_{SC}$. Due to the interaction of the atoms with the radiation field of the cavity, which only changes the average number of photons by a small amount, an interference of the "classical states"  
corresponding to the quantum coherent states $\rv{z}$ and $\rv{-z}$ is, in addition, induced. In this sense we have therefore obtained a novel quantum-mechanical interferometer which is experimentally feasible.  
In our numerical examples we have used physical parameters from  
some recent experiments \cite{brune&96,benson&97}.  
but extrapolated the atom transit time $t$   
to rather large values of $gt$. It is, of course,  an experimental  
 challenge to obtain a one-atom source and   
atomic life-times of the atomic states involved  
 such that these large values of $gt$ can be reached.  
\vspace{3mm}  
\newpage  
\begin{center}  
{\Large \bf ACKNOWLEDGEMENT}  
\end{center}  
%  
%\vspace{3mm}  
%  
The authors wish to thank A. De R\'{u}jula and the members of the  
TH-division at CERN for the warm hospitality while the present work  
was completed.  B.-S. S. wish to thank H. Walther for discussions and  
for providing a guide to the experimental work and for discussions.  
The research has been supported in part by the Research Council of  
Norway under contract no. 118948/410.  
 \vspace{3mm}  
%  
%  
%\newpage  
%  
%  
  
\appendix  
  
\begin{center}  
{\Large \bf APPENDIX}  
\end{center}  
\newcommand{\seqnoll}{\setcounter{equation}{0}}  
\renewcommand{\theequation}{A.\arabic{equation}}  
  
%\numberwithin{equation}{section}  
%{\hspace{-0.7cm}\Huge \bf Appendix}  
%\section{Some important relations}  
\seqnoll  
%\label{app:eqmo}  
  
In order to derive the equations of motion in a convenient form, we notice  the following useful relations  
 \begin{equation}  
  a^{*} \ket{\psi^{\pm}_{n}}  =     
\quad\frac{1}{2}\left( \sqrt{n+1}\pm  
    \sqrt{n+2}\right) \ket{\psi^{+}_{n+1}}   
   +  \quad\frac{1}{2}\left( \sqrt{n+1}\mp  
    \sqrt{n+2}\right) \ket{\psi^{-}_{n+1}} \label{eq:apsi} \ ,  
 \end{equation}   
%\\[5mm]  
and  
 \begin{equation}  
  a \ket{\psi^{\pm}_{n}} = \quad\frac{1}{2}\left( \sqrt{n}\pm  
    \sqrt{n+1}\right) \ket{\psi^{+}_{n-1}}   
  + \quad  \frac{1}{2}\left( \sqrt{n}\mp  
    \sqrt{n+1}\right) \ket{\psi^{-}_{n-1}} \ ,  
 \end{equation}  
from which we derive  
 \begin{equation}  
  a^{*}a \ket{\psi^{\pm}_{n}}  = \left(n+\frac{1}{2}\right)  
  \ket{\psi^{\pm}_{n}} -   \frac{1}{2} \ket{\psi^{\mp}_{n}} \ .  
  \label{eq:akorspsi}  
 \end{equation}  
%  
%\section{Equations of motion}  
  
%  
  
\noindent Using the definition Eq.(\ref{Wdef}) we then find the following exact differential  
equations of the diagonal elements of Eq.(\ref{eq:Damping})  
 \begin{eqnarray}  
    &&\hspace{2cm}\bra{\psi^{\pm}_{n}}\dot{W}(t) \ket{\psi^{\pm}_{n}}  
 =   
  2\kappa (n_{b}  +1) \bigpar{\{ }      
  \Gamma_{+ , n+1} \wpp{n+1}  \nonumber \\   
    &&\hspace{2cm}+ \Gamma_{-,n+1}\wmm{n+1}  
  - (n+\frac{1}{2}) \wpp{n} \nonumber \\  
        &&+ \frac{1}{4}\bigpar{[} \ex{-2}{n+1} \wspm{n}   
     -\ex{-2}{n+2}\wspm{n+1} + c.c. \bigpar{]}  \bigpar{\}}\nonumber \\  
      &&+ 2 \kappa n_b  \bigpar{\{ }     
      \Gamma_{+ , n} \wpp{n-1}    
     + \Gamma_{-,n}\wmm{n-1}  
     - (n+\frac{3}{2})\wpp{n} \nonumber\\  
   &&+  \frac{1}{4}\bigpar{[}\ex{-2}{n+1} \wspm{n}     
     - \ex{-2}{n}\wspm{n-1} +c.c.  \bigpar{]} \bigpar{\}} \ ,  
      %\hspace{1cm}     
\label{eq:eqmo1}  
%  \end{split}  
%\end{gather}  
 \end{eqnarray}  
for $n\geq 1$ and  
 \begin{eqnarray}  
&& \bra{\psi^{\pm}_{0}}  \dot{W}(t) \ket{\psi^{\pm}_{0}}  =  
      2\kappa (n_{b}   
      +1) \bigpar{\{ } \Gamma_{+,1} \wpp{1}    
      +\Gamma_{-,1}\wmm{1} \nonumber \\  
&& - \frac{1}{2}\wpp{0}   
      +  \frac{1}{4}\bigpar{[} e^{-2igt} \wspm{0}  
     - \ex{-2}{2}\wspm{1} +c.c. \bigpar{]} \bigpar{\}}  
      \nonumber \\  
&&+ 2\kappa n_b  \bigpar{\{ }     
        \bra{\psi_0}W(t)\ket{\psi_0} - 3\wpp{0}   
     + \frac{1}{2} \bigpar{[} e^{-2igt}  \wspm{0} +  
    c.c. \bigpar{]}  \bigpar{\}} . \nonumber \\  
\label{eq:eqmo2}  
 \end{eqnarray}  
A dot denotes differentiation with respect to the time variable $t$.  
With the same procedure as above, we also find for $n\geq 1$ the equation  
 \begin{eqnarray}  
&& \hspace{2cm} \bra{\psi^{\pm}_{n}} \dot{W}(t)\ket{\psi^{\mp}_{n}}   = -  
    2\kappa (n_{b}   
    +1) \bigpar{\{ }   (n+\frac{1}{2})\wpm{n}\nonumber \\  
 &&- \Gamma_{+ , n+1} \exx{\pm 2}{n+1}{-}{n+2}\wpm{n+1}  
- \Gamma_{-,n+1} \exx{\pm 2}{n+1}{+}{n+2} \wmp{n+1}\nonumber \\  
&& - \frac{1}{4}\ex{\pm 2}{n+1}\bigpar{[}\wsmm{n} + \wspp{n}  
       - \wsmm{n+1} - \wspp{n+1} \bigpar{]}\bigpar{\}}\nonumber \\  
&&- 2\kappa n_b  \bigpar{\{ }     
    (n+\frac{3}{2})\wpm{n}  
       - \Gamma_{+ , n} \exx{\pm 2}{n+1}{-}{n}\wpm{n-1}\nonumber\\  
&&- \Gamma_{-,n} \exx{\pm 2}{n+1}{+}{n} \wmp{n-1}  
       - \frac{1}{4}\ex{\pm 2}{n+1}\bigpar{[}\wsmm{n} + \wspp{n} \nonumber\\  
&&\hspace{3cm}- \wsmm{n-1} - \wspp{n-1} \bigpar{]}\bigpar{\}} \ ,   
\label{eq:eqmo3}  
 \end{eqnarray}  
as well as  
 \begin{eqnarray}  
%  \begin{split}  
     \bra{\psi^{\pm}_{0}} \dot{W}(t)\ket{\psi^{\mp}_{0}}   =   
    2\kappa (n_{b}   
    +1) \bigpar{\{ }    -\frac{1}{2}\wpm{0}    
     + \Gamma_{+ , 1} e^{\pm 2igt (1-\sqrt{2})}  \wpm{1}  \nonumber \\  
      + \Gamma_{-,1}  e^{\pm 2igt (1+\sqrt{2})} \wmp{1}  
      + \frac{1}{4} e^{\pm 2igt} \bigpar{[}\wsmm{0} + \wspp{0}  
    \nonumber \hspace{1.5cm} \\  
      - \wsmm{1} - \wspp{1} \bigpar{]}\bigpar{\}}   
      + 2\kappa n_b  \bigpar{\{ }     
    - \frac{3}{2}\wpm{0} - \frac{1}{2} \bra{\psi_0}W(t)\ket{\psi_0}  
    \nonumber \\   
      + \frac{1}{4} e^{\pm 2igt}   \bigpar{[}  \wspp{0} +  
    \wsmm{0} \bigpar{]}\bigpar{\}}  \ . \hspace{3cm}\label{eq:eqmo4}  
 \end{eqnarray}  

\noindent Here we have made use of the notation  
\begin{equation}  
\Gamma_{\pm , n} = (\sqrt{n+1} \pm  \sqrt{n})^2/4 \ .  
\end{equation}  
With regard to the time-scale $1/\kappa$ of cavity damping, the exponential functions in  
Eqs.(\ref{eq:eqmo1})-(\ref{eq:eqmo2}) and Eqs.(\ref{eq:eqmo3})-(\ref{eq:eqmo4}) will vary rapidly and can therefore be neglected provided $\kappa \ll g$, which we assume to be valid.  
%  
%   
%==================== bibliography===========================  
%   
     
%   
%%%%%%%%%%%%%%%%%%%%%%%%%%%%%%%%%%%%%%%%%%%%%%%%%%%%%%%%%%%%%%%%%%  
\end{document}